\documentclass[preprint,12pt, num-names]{elsarticle}

\usepackage{setspace}
\usepackage{hyperref}
\usepackage{graphicx}
\usepackage{subfigure}
\usepackage{mhchem}
\usepackage{float}
\usepackage{booktabs}
\usepackage{amsmath}
\usepackage{algorithm}
\usepackage{algorithmic}
\usepackage{amsthm}
\usepackage{threeparttable}
\usepackage{enumitem}
\usepackage{multirow}
\usepackage{longtable}
\usepackage{chemformula}
\usepackage{siunitx}
\usepackage{hyperref}

\usepackage{amssymb}
\usepackage{amsmath}

\usepackage{lineno}

\journal{Combustion and Flame}

\begin{document}

\setstretch{1.0}

\begin{frontmatter}



\title{Analysis of Information Loss on Composition Measurement in Stiff Chemically Reacting Systems}


\author[1]{Yiming Lu} 
\author[1]{Xu Zhu}
\author[1]{Long Zhang}
\author[1]{Hua Zhou\corref{cor1}}
%

\affiliation[1]{organization={Institute for Aero Engine},
	addressline={Tsinghua University}, 
	city={Beijing},
	postcode={100084}, 
	state={},
	country={China}}

%

\cortext[cor1]{Corresponding author. Email: zhouhua@tsinghua.edu.cn}

\begin{abstract}

Gas sampling methods have been crucial for the advancement of combustion science, enabling analysis of reaction kinetics and pollutant formation. However, the measured composition can deviate from the true one because of the potential residual reactions in the sampling probes. This study formulates the initial composition estimation in stiff chemically reacting systems as a Bayesian inference problem, solved using the No-U-Turn Sampler (NUTS). Information loss arises from the restriction of system dynamics by low dimensional attracting manifold, where constrained evolution causes initial perturbations to decay or vanish in fast eigen-directions in composition space. This study systematically investigates the initial value inference in combustion systems and successfully validates the methodological framework in the Robertson toy system and hydrogen autoignition. Furthermore, a gas sample collected from a one-dimensional hydrogen diffusion flame is analyzed to investigate the effect of frozen temperature on information loss.
The research highlights the importance of species covariance information from observations in improving estimation accuracy and identifies how the rank reduction in the sensitivity matrix leads to inference failures. Critical failure times for species inference in the Robertson and hydrogen autoignition systems are analyzed, providing insights into the limits of inference reliability and its physical significance.

\end{abstract}




\end{frontmatter}


\newpage

\section{Introduction}
\label{sec:intro}

Combustion diagnostic methods, such as probe sampling \cite{nasr1984,mitani1999analyses,chen2010gas}, have significantly advanced combustion science, facilitating the analysis of reaction kinetics and pollutant formation. The oxidation of hydrocarbon fuels typically involves a large number of chemical species and spans a wide range of timescales \cite{hucknall2012chemistry,westbrook1984chemical}, making species measurement inherently challenging.

According to the second law of thermodynamics, entropy in an isolated reactive system increases as chemical reactions progress. While the final products can often be identified, the details of intermediate states may be irretrievably lost.
For instance, with probe sampling, the initial state of the sample gas mixture becomes less accessible as the mixture evolves due to potential residual reactions. As a result, the composition of the collected gas mixture can differ significantly from the initial composition, leading to substantial measurement errors. 
These errors can be effectively reduced using cooling or dilution devices \cite{Hayakawa2020Novel,Kong2013Emission,MITANI19962917,larosa2010}, which lower the temperature or concentration of the sampled gas mixture to a level that freezes chemical reactions.  
While these efforts primarily focus on the effects of quenching rates on compostition preservation, they often overlook the statistical analysis of error propagation during the quenching process.

In addition, even with a sufficiently fast quenching rate, measurement errors can still arise from residual reactions during the quenching process. Currently, no theoretical analysis has been conducted to explore this issue from the intrinsic nature of the governing equations.
One possible approach to reconstruct the initial gas state prior to residual reactions is to formulate an inverse problem based on the known current state. With diffusion and boundary layer effect \cite{Yanagi1972Effect,Jiang2007Flow,Zhu2016Computational} being neglected, the probe can be modeled as an ordinary differential equation (ODE) and then applying reverse integration, the initial state could, in principle, be recovered. However, the inverse interpretation of stiff systems is practically challenging. Moreover, this approach neglects measurement errors in the current state.
A more suitable approach would be to infer the mean and standard deviation of the initial state, given the mean and standard deviation of the current state. This accounts for uncertainties in the measurements, leading to a more robust estimate of the initial gas composition.

This study conducts a theoretical analysis of the factors influencing the estimation of the initial composition, based on the intrinsic properties of the governing equations. The representative chemically reacting system relevant to the gas sample is modeled using a set of time-dependent ordinary differential equations (ODEs).  
Inspired by previous studies on uncertainty minimization in kinetic parameter estimation \cite{sander2011developing,kastner2013bayesian,mosbach2014bayesian,su2023kinetics}, the initial composition estimation problem is formulated as a Bayesian inference problem. The mean and standard deviation of the current state define a normal likelihood function at the observation time, while an uninformative prior on the initial composition is specified as a truncated uniform distribution. The posterior distribution is then obtained as the product of the prior and the likelihood.  
To sample from this posterior distribution, the No-U-Turn Sampler (NUTS) \cite{Hoffman2014} is employed. NUTS is advantageous due to its ease of use, requiring only the tuning of the target acceptance rate $\delta_\mathrm{acc}$. Additionally, it provides stable sampling efficiency for high-dimensional probability distributions, making it more suitable than traditional random-walk Markov Chain Monte Carlo (MCMC) methods for solving the high-dimensional initial composition estimation problem in chemical reaction systems.  
The proposed methodology is demonstrated in both the Robertson toy system \cite{robertson1966reaction} and the hydrogen autoignition system that exhibit chemical stiffness.

This work is the first to comprehensively investigate initial composition estimation as a Bayesian inference problem and to solve it using the No-U-Turn Sampler (NUTS). The remainder of this paper is structured as follows.  
In Section \ref{sec:metho}, the Bayesian inference framework for initial composition estimation is introduced, along with an analysis of inference failures. The NUTS algorithm is implemented using the adjoint sensitivity method.  
Section \ref{sec:RaD} examines inference performance at different observation times for both the Robertson and hydrogen/oxygen systems, identifying instances of inference failure and determining the critical failure time ranges for different species. A theoretical connection is established between intrinsic low-dimensional attracting manifolds and inference failures, with the reduction in the rank of the sensitivity matrix identified as a key indicator of information loss over time.  
Finally, conclusions are presented in Section \ref{sec:Ccl}.  

\newpage

\section{Methodology}
\label{sec:metho}

\subsection{ODEs for stiff chemical systems}

A homogeneous reactive system with $ n_s $ species, $ n_e $ elements, and $ n_r $ reactions evolves according to:
\begin{equation}
	\label{eq:ODE_formula}
        \begin{aligned}
        \frac{d}{dt} (\rho\phi_i) &= {W_i} \sum_{r=1}^{n_r} (\nu_{i,r}'' - \nu_{i,r}') R_r, \quad i = 1, \dots, n_s, \\
        \frac{d}{dt} (\rho c_pT) &= - \sum_{r=1}^{n_r} R_r \sum_{i=1}^{n_s} h_i {W_i} (\nu_{i,r}'' - \nu_{i,r}') - q_{\text{loss}},
     \end{aligned}
\end{equation}
or in vector form, $ \frac{d}{dt} \boldsymbol{\phi} = \boldsymbol{S}(\boldsymbol{\phi}) $, where $ \boldsymbol{\phi} = [\phi_1, \dots, \phi_{n_s}]^T $ represents species mass fractions. Here, $ W_i $ is the molar mass, $ \rho $ the mixture density, $ R_r $ the reaction rate, and $ \nu_{i,r}' $, $ \nu_{i,r}'' $ the stoichiometric coefficients. The energy equation includes the specific heat $ c_p $ and enthalpy $ h_i $.
For an adiabatic, isobaric system, $ q_{\text{loss}} = 0 $, and the enthalpy $ h $ and pressure $ P $ remain constant.
Assuming negligible effects from the probe’s internal walls on the flow, mapping the distance traveled within the probe to the time variable \(t\) and accounting for heat loss \(q_{\text{loss}}\) yields a similar ODE system for the composition evolution.

To emphasize the role of the initial value $\boldsymbol{\phi}_{0}$ in integrating the ODE system, Equation~\eqref{eq:ODE_formula} can be reformulated as
\begin{equation}
	\label{eq:ODE_formula_remap_format}
	\frac{\partial}{\partial t} \boldsymbol{R}(\boldsymbol{\phi}_{0}, t) = \boldsymbol{S}(\boldsymbol{R}(\boldsymbol{\phi}_{0}, t)),
\end{equation}
where $\boldsymbol{R}(\boldsymbol{\phi}_0, t)$ represents the reaction map \cite{ren2006geometry,ren2006use}, and the solution to the system is given by $\boldsymbol{\phi}(t) = \boldsymbol{R}(\boldsymbol{\phi}_0, t)$ with the initial condition $\boldsymbol{\phi}_0$.  
The dynamics of a stiff reactive system are governed by the Jacobian matrix:
$\boldsymbol{J}(\boldsymbol{\phi}) = {\partial \boldsymbol{S}(\boldsymbol{\phi})}/{\partial \boldsymbol{\phi}^T}$.
Its eigenvalues span a wide range \cite{Razon1987Multiplicities,Lu2009Toward}, reflecting disparate chemical timescales.A small perturbation \(\delta \boldsymbol{\phi}\) evolves according to
\begin{equation}
\frac{d}{dt} \delta \boldsymbol{\phi} = \boldsymbol{J}(\boldsymbol{\phi}) \delta \boldsymbol{\phi}.
\end{equation}
Assuming the eigen-directions of \(\boldsymbol{J}\) remain unchanged, we can perform an eigen-decomposition (following \citet{maas1992simplifying}) to obtain
\begin{equation}
\frac{d}{dt} \delta \boldsymbol{\psi} = \boldsymbol{\Lambda} \delta \boldsymbol{\psi}.
\end{equation}
where $ \delta \boldsymbol{\psi} = \boldsymbol{L} \delta \boldsymbol{\phi} $, with $ \boldsymbol{L} $ as the left eigenmatrix of $ \boldsymbol{J} $, and $ \boldsymbol{\Lambda} = \mathrm{diag}(\lambda_1, \dots, \lambda_{n_s}) $, ordered such that $ |\mathrm{Re}(\lambda_1)| \ge \dots \ge |\mathrm{Re}(\lambda_{n_s})| $.
Element conservation ensures $ \lambda_{n_s - n_e + 1}, \dots, \lambda_{n_s} = 0 $. The remaining $ (n_s - n_e) $ eigenvalues are maily negative, leading to timescale separation ${|\mathrm{Re}(\lambda_1)|}/{|\mathrm{Re}(\lambda_{n_s - n_e})|} \gg 1$ \cite{pope2013small}.This enforces rapid decay of fast modes, constraining trajectories to a low-dimensional manifold where slow dynamics dominate.

\begin{figure}[htbp]
	\centering
	\includegraphics[width=0.5\linewidth]{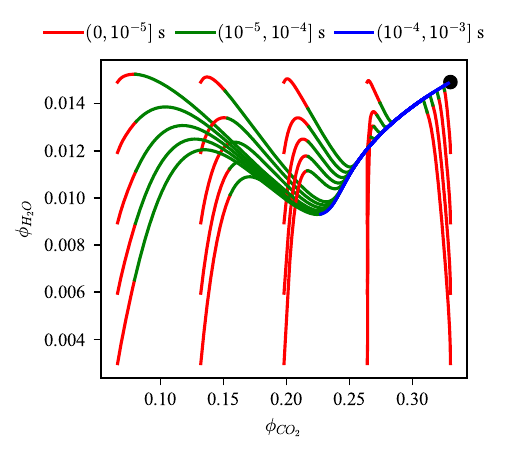}
	\caption[Chemical reaction trajectories towards the attracting manifold]{Chemical reaction trajectories in the mole fraction phase plane of \ch{CO2}-\ch{H2O}, simulated under GRI-Mech 3.0 \cite{GRIMech3.0}. The system was initialized with 25 different conditions while maintaining the same elemental ratios, pressure, and enthalpy. The trajectories rapidly converge within $10^{-4}$ seconds to the equilibrium state (black dots).}
	\label{fig:trajectories}
\end{figure}

The properties of chemical attracting manifolds have been extensively studied in \cite{maas1992simplifying,maas2020time_reversal,ren2006geometry,ren_species_2005,ren2006use,ren_kinetics-based_2017}. Figure~\ref{fig:trajectories} illustrates that the trajectories of a hydrocarbon system, initialized from different states, converge toward the attracting manifold and eventually adhere to it. This behavior facilitates the characterization of reactive systems with a few major species, where the reconstruction of full thermodynamic states can be achieved using a few major species or minimal principal components (e.g., mixture fraction, reaction progress variable). In addition to common dimensionality reduction methods such as Principal Component Analysis (PCA) \cite{mirgolbabaei2014nonlinear,Mirgolbabaei2014,Bilgari2012,Parente2011,Sutherland2009}, low-dimensional manifold methods can reduce the dimensionality of state variables in combustion simulations. Prominent low dimensional manifold methods include: Quasi-Steady State Assumption (QSSA) \cite{smooke_reduced_1991}, Partial Equilibrium Assumptions (PEA) \cite{ramshaw1980partial,rein1992partial}, Rate-Controlled Constrained Equilibrium (RCCE) \cite{keck_rate-controlled_1971,keck_rate-controlled_1990,ren_kinetics-based_2017}, Empirical Low-Dimensional Manifolds (ELDM) \cite{yang2013empirical,ren_Reduced,ren2007transport,Pope2004} Intrinsic Low-Dimensional Manifolds (ILDM) \cite{maas1992simplifying}, ICE-PIC \cite{ren_species_2005,ren2006geometry,ren2006use,ren2007application,ren2006invariant}, Computational Singular Perturbation (CSP) \cite{lam1993using,lam1994csp}, Relaxation Redistribution Method (RRM) \cite{chiavazzo2011adaptive,chiavazzo2012approximation,kooshkbaghi2014global}, and the Equation-Free approach \cite{kevrekidis2004equation,chiavazzo2014reduced}. All of these methods leverage the behavior of attracting manifolds, and this concept has significantly enhanced the analysis of the sensitivity matrix in relation to convergence behavior. 

To quantify the convergence behavior of reaction trajectories, \citet{ren2006geometry} introduced the sensitivity matrix,
\begin{equation}
    \boldsymbol{A} = \frac{\partial \boldsymbol{R}(\boldsymbol{\phi}_0, t)}{\partial \boldsymbol{\phi}_0^T},
\end{equation}
which is computed using Algorithm~\ref{alg:forward_sensitivity}. The sensitivity matrix describes the evolution of an initial perturbation $d \boldsymbol{\phi}_0 \in \mathbb{R}^{n_s}$ along the reaction trajectory:
\begin{equation}
    d \boldsymbol{R} = \boldsymbol{A} d \boldsymbol{\phi}_0.
    \label{eq:dphi}
\end{equation}
This transformation maps an initial infinitesimal perturbation ball,
\begin{equation}
    \{ \boldsymbol{\phi} = \boldsymbol{\phi}_0 + d\boldsymbol{\phi}_0 \mid \| d\boldsymbol{\phi}_0 \| \leq dr \},
\end{equation}
into a hyper-ellipsoid,
\begin{equation}
    \{ \boldsymbol{\phi} = \boldsymbol{R} + d\boldsymbol{R} \mid \| \boldsymbol{A}^{-1} d\boldsymbol{R} \| \leq dr \},
\end{equation}
whose principal axes are determined by the sensitivity matrix.  
A detailed analysis of the relationship between the sensitivity matrix, information loss, and inference failure will be provided in Sections~\ref{sec:mappcov} and~\ref{sec:species_times}.

\begin{algorithm}[htpb]
	\caption{Forward Sensitivity Method}
	\label{alg:forward_sensitivity}
	\begin{algorithmic}[1]
		\REQUIRE Initial condition $\boldsymbol{\phi}_0$; time range $[0, t_\text{end}]$; source term $\boldsymbol{S}(\cdot)$
		\ENSURE State variable $\boldsymbol{\phi}(t)$ and sensitivity matrix $\boldsymbol{A}(t)$ for $t \in [0, t_\text{end}]$
		\STATE Initialize $\boldsymbol{\phi}(0) \gets \boldsymbol{\phi}_0$, $\boldsymbol{A}(0) \gets \boldsymbol{I}$ (identity matrix)
		\STATE Define Jacobian matrix $\boldsymbol{J}(\boldsymbol{\phi}) \gets \partial \boldsymbol{S}(\boldsymbol{\phi}) / \partial \boldsymbol{\phi}^T$
		\STATE Simultaneously integrate the following equations for $t \in [0, t_\text{end}]$:
		\STATE $\quad \frac{d\boldsymbol{\phi}}{dt} = \boldsymbol{S}(\boldsymbol{\phi}) \quad \text{with initial condition } \boldsymbol{\phi}(0) = \boldsymbol{\phi}_0$
		\STATE $\quad \frac{d\boldsymbol{A}}{dt} = \boldsymbol{J}(\boldsymbol{\phi}) \boldsymbol{A} \quad \text{with initial condition } \boldsymbol{A}(0) = \boldsymbol{I}$
		\RETURN $\boldsymbol{\phi}(t)$, $\boldsymbol{A}(t)$
	\end{algorithmic}
\end{algorithm}

\newpage

\subsection{Bayesian inference for initial composition}

In this study, we address the problem of inferring the likely initial composition distribution from the observed mass-fraction distribution at a given observation time and comparing it to the underlying ground truth distribution that generates the observed data.

\begin{figure}[htbp]
	\centering
	\includegraphics[width=1.0\linewidth]{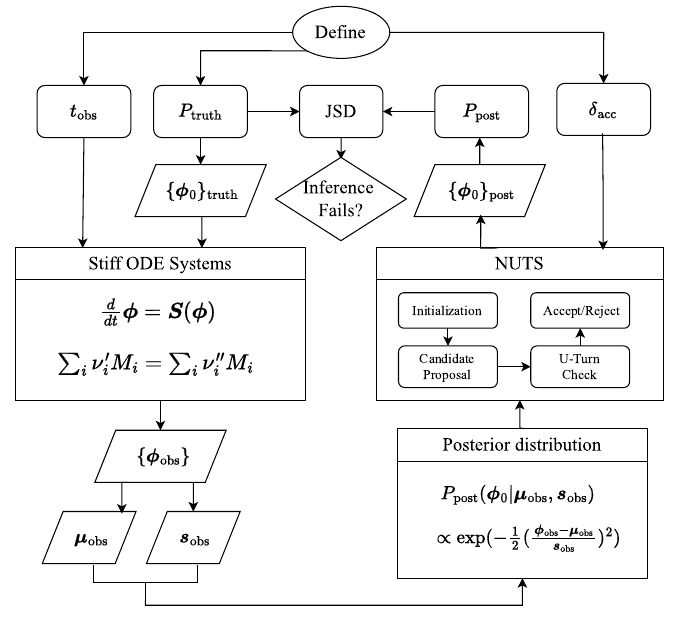}
	\caption[Research approach for initial composition estimation using NUTS]{Research approach for initial composition estimation using NUTS. The process begins by defining the ground truth distribution of the initial composition and obtaining a set of samples $\{\boldsymbol{\phi}_0\}_{\mathrm{truth}}$ from this distribution. These samples are then input into the reaction system to compute statistical quantities, such as the mean $\boldsymbol{\mu}_\mathrm{obs}$ and standard deviation $\boldsymbol{s}_\mathrm{obs}$, at a specific observation time $t_{\mathrm{obs}}$. Using Bayes' theorem, the posterior probability density function $P_\mathrm{post}(\boldsymbol{\phi}_0)$ of the initial parameter values is constructed. NUTS is then employed to draw samples from this distribution, and the resulting posterior samples are compared to the ground truth to validate the accuracy of the inference results.}
	\label{fig:process_research}
\end{figure}
 
As illustrated in Figure~\ref{fig:process_research}, the ground truth distribution $ P_{\mathrm{truth}}(\boldsymbol{\phi}_0) $ is defined as a truncated Gaussian with mean $ \boldsymbol{\mu}_{0} $, standard deviation $ \boldsymbol{s}_{0} $, and bounds $ \boldsymbol{\phi}_{0,\mathrm{upper}} $ and $ \boldsymbol{\phi}_{0,\mathrm{lower}} $. A set of $ n_{\mathrm{truth}} $ initial samples, $ \{\boldsymbol{\phi}_0\}_{\mathrm{truth}} $, is drawn from this distribution. These states are then integrated through the ODE system up to the observation time $ t_{\mathrm{obs}} $, yielding the observation samples $ \{\boldsymbol{\phi}_{\mathrm{obs}}\}_{\mathrm{truth}} $. A statistical analysis provides the mean $ \boldsymbol{\mu}_{\mathrm{obs}} $ and standard deviation $ \boldsymbol{s}_{\mathrm{obs}} $. 

The likelihood function is assumed to be a multivariate Gaussian:
\begin{equation}
	\mathcal{L}(\boldsymbol{\phi}(t_{\mathrm{obs}});\boldsymbol{\mu}_{\mathrm{obs}},\boldsymbol{s}_{\mathrm{obs}}) =
	\frac{1}{(2\pi)^{\frac{n_{s}}{2}} \prod_{i=1}^{n_{s}} s_{\mathrm{obs},i}}
	\exp \left( -\frac{1}{2} \sum_{i=1}^{n_{s}} \left(\frac{\phi_{i}(t_{\mathrm{obs}}) - \mu_{\mathrm{obs},i}}{s_{\mathrm{obs},i}}\right)^2 \right),
\end{equation}
where $ \boldsymbol{\phi}(t_{\mathrm{obs}}) $ is the system state at $ t_{\mathrm{obs}} $. The prior is a uniform distribution $ P_{\mathrm{prior}}(\boldsymbol{\phi}_0) \propto 1 $, defined over the domain $ K^{n_s} $, bounded by $ \boldsymbol{\phi}_{0,\mathrm{upper}} $ and $ \boldsymbol{\phi}_{0,\mathrm{lower}} $. The posterior distribution is then:
\begin{equation}
	\label{eq:posterior}
	\begin{aligned}
		P_{\mathrm{post}}(\boldsymbol{\phi}_0 | \boldsymbol{\mu}_{\mathrm{obs}}, \boldsymbol{s}_{\mathrm{obs}}) &=
		\frac{\mathcal{L}(\boldsymbol{\phi}(t_{\mathrm{obs}}); \boldsymbol{\mu}_{\mathrm{obs}}, \boldsymbol{s}_{\mathrm{obs}}) P_{\mathrm{prior}}(\boldsymbol{\phi}_0)}
		{\int_{K^{n_s}} \mathcal{L}(\boldsymbol{\phi}(t_{\mathrm{obs}}); \boldsymbol{\mu}_{\mathrm{obs}}, \boldsymbol{s}_{\mathrm{obs}}) P_{\mathrm{prior}}(\boldsymbol{\phi}_0) d\boldsymbol{\phi}_0} \\
		& \propto \exp \left( -\frac{1}{2} \sum_{i=1}^{n_s} \left( \frac{\phi_i(t_{\mathrm{obs}}) - \mu_{\mathrm{obs},i}}{s_{\mathrm{obs},i}} \right)^2 \right).
	\end{aligned}
\end{equation}
Since $ \boldsymbol{\phi}(t_{\text{obs}}) $ lacks an analytical solution, and the integral in Equation~\eqref{eq:posterior} is intractable, we approximate the posterior using NUTS, generating samples $ \{\boldsymbol{\phi}_0\}_{\mathrm{post}} $. If $ \{\boldsymbol{\phi}_0\}_{\mathrm{post}} $ closely matches $ \{\boldsymbol{\phi}_0\}_{\mathrm{truth}} $, the inference is effective. Both $ \{\boldsymbol{\phi}_0\}_{\mathrm{post}} $ and $ \{\boldsymbol{\phi}_0\}_{\mathrm{truth}} $ share the same initial enthalpy $ h_0 $ and pressure $ P_0 $.

The similarity between the posterior and ground truth distributions is quantified using the Jensen-Shannon distance (JSD) \cite{Lin1991Divergence}:
\begin{equation}
	\label{eq:JSD}
	D_\mathrm{JS}(P_\mathrm{truth} \parallel P_\mathrm{post}) =
	\sqrt{\frac{1}{2} D_\mathrm{KL}(P_\mathrm{truth} \parallel M) + \frac{1}{2} D_\mathrm{KL}(P_\mathrm{post} \parallel M)},
\end{equation}
where
\begin{equation}
    M = \frac{1}{2} (P_\mathrm{truth} + P_\mathrm{post}),
\end{equation}
and $D_\mathrm{KL}$ denotes the Kullback-Leibler (K-L) divergence, defined as:
\begin{equation}
	\label{eq:KLD}
	D_\mathrm{KL}(P \parallel Q) =
	\int_{\mathcal{D}} P(x) \log_{10} \frac{P(x)}{Q(x)} \,dx.
\end{equation}
Here, $\mathcal{D} = \{ x \mid P(x) Q(x) \neq 0 \}$ represents the domain where both distributions are nonzero.  
The JSD is a symmetrized and smoothed version of the K-L divergence, providing a robust measure of similarity between two distributions. Its value ranges from 0 to 1, where smaller values indicate greater similarity between the posterior and ground truth distributions, signifying higher inference accuracy.

The probabilistic inference in this work utilizes Pyro \cite{bingham2019pyro} in Python and Turing.jl \cite{ge2018turing} in Julia, both providing NUTS for Bayesian sampling. ODE integration and adjoint sensitivity analysis are handled by \href{https://github.com/patrick-kidger/diffrax}{Diffrax} library in Python and DifferentialEquations.jl \cite{rackauckas2017differentialequations} with \href{https://docs.sciml.ai/SciMLSensitivity/stable/}{SciMLSensitivity.jl} in Julia. The reaction mechanism for the Hydrogen/Oxygen system is parsed using \href{https://deng-mit.github.io/Arrhenius.jl/dev/}{Arrhenius.jl} \cite{ji2022sgd}.

\newpage

\section{Results and Discussion}
\label{sec:RaD}

An eigen-decomposition of the Robertson toy system is performed to analyze fast and slow reaction directions, followed by initial composition inference with and without observed covariance. A critical inference failure time emerges, linked to covariance inclusion and sensitivity matrix singularity. The hydrogen autoignition system is then used to validate these findings by decomposing its sensitivity matrix into Conserved and Reaction Spaces, preserving information only in the former. Finally, under hydrogen diffusion flame sampling conditions, we observe the rank of the sensitivity matrix to demonstrate how cooling can mitigate the loss of information.

\subsection{Robertson system ($n_e=1$, $n_s = n_r = 3$)}

\begin{figure}[htpb]
	\centering
	\includegraphics[width=0.5\linewidth]{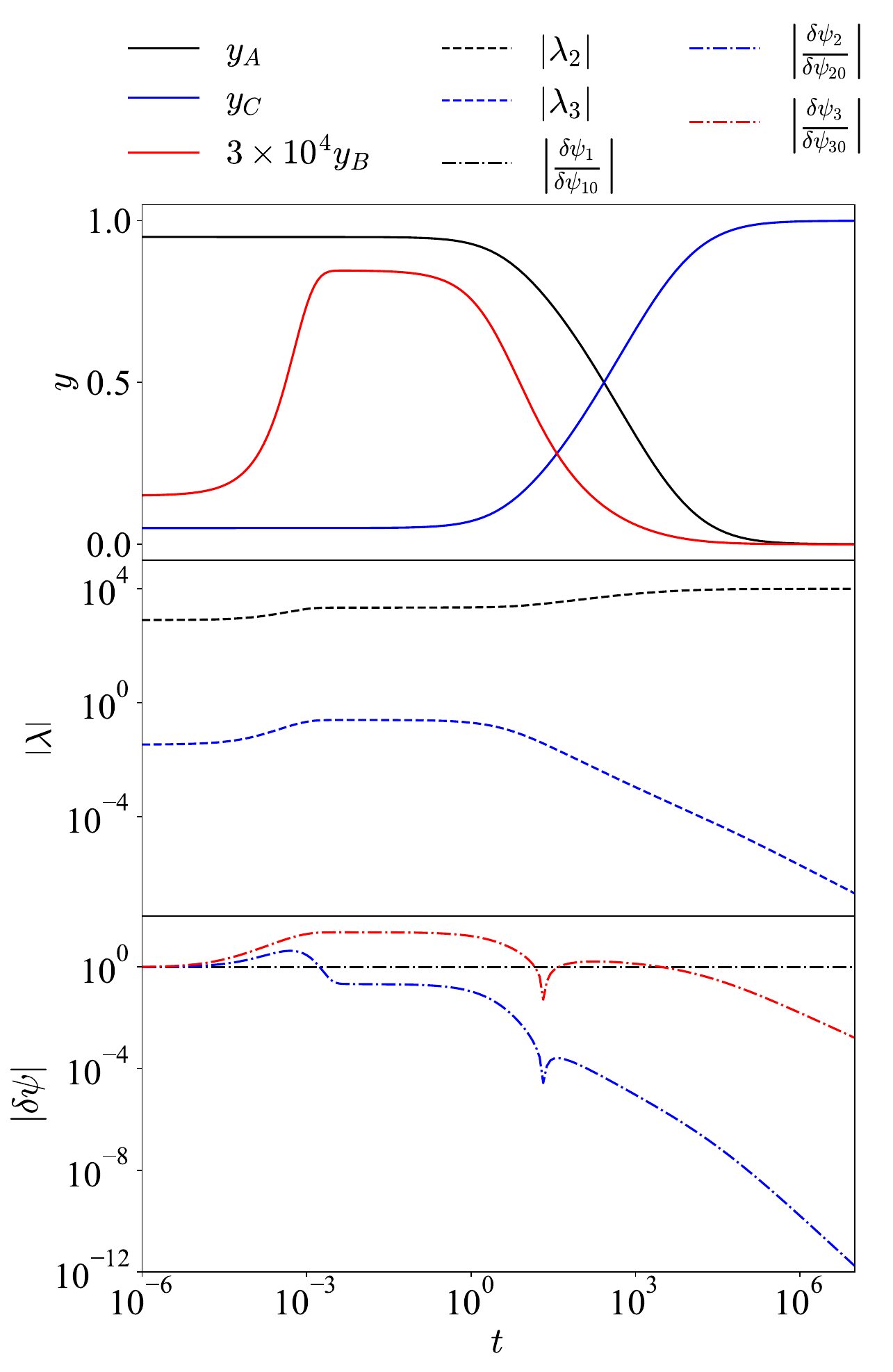}
	\caption{Concentration profiles, eigenvalues, and decoupled perturbations as functions of time for the Robertson system with given initial values and rate constants. The initial perturbation is set as $\delta\boldsymbol{y}_0 = [10^{-3},10^{-7},10^{-3}]$.}
	\label{fig:robertson}
\end{figure}

The Robertson system consists of the following three reactions:
\begin{equation}
    \begin{array}{c}
        A \stackrel{k_1}{\longrightarrow} B,\\
        B + B \stackrel{k_2}{\longrightarrow} C + B,\\
        B + C \stackrel{k_3}{\longrightarrow} A + C.
    \end{array}
\end{equation}
The reaction rate constants are given by $[k_1, k_2, k_3] = [0.04, 3 \times 10^7, 10^4]$.
The initial conditions are denoted as $\boldsymbol{y}_0 = [y_{A0}, y_{B0}, y_{C0}]$, where $y_A$, $y_B$, and $y_C$ represent the concentrations of species $A$, $B$, and $C$, respectively.  
The evolution of the species concentrations is governed by the following system of ordinary differential equations:
\begin{equation}
	\label{eq:rober_rhs}
	\begin{aligned}
		\frac{dy_{A}}{dt} &= -k_{1} y_{A} + k_{3} y_{B} y_{C}, \\
		\frac{dy_{B}}{dt} &= k_{1} y_{A} - k_{2} y_{B}^{2} - k_{3} y_{B} y_{C}, \\
		\frac{dy_{C}}{dt} &= k_{2} y_{B}^{2}.
	\end{aligned}
\end{equation}  
Using representative initial concentrations of $\boldsymbol{y}_0 = [0.95, 5 \times 10^{-6}, 0.05]$, the corresponding concentration evolutions are shown in Figure~\ref{fig:robertson}. Species $A$ and $C$ act as the primary reactant and product, respectively, and their concentrations remain on a comparable scale. In contrast, the concentration of $B$ is significantly lower, approximately $10^{-5}$ times that of $A$ and $C$.  

The eigenvalues of the Jacobian matrix are given by:
\begin{equation}
		\lambda_1 = 0,~
		\lambda_2 = \frac{-E - Q}{2} < 0,~
		\lambda_3 = \frac{-E + Q}{2} < 0,
\end{equation}
where $E = k_1 + k_3 y_C + 2 k_2 y_B$, $F = 2 k_1 k_2 y_B + 2 k_2 k_3 y_B^2$, $Q = \sqrt{E^2 - 4F}.$

The left eigenmatrix is given by:
\begin{equation}
	\boldsymbol{L} =
	\begin{bmatrix}
		1 & 1 & 1 \\
		Q_1 - \frac{1}{2} & Q_1 - Q_2 - \frac{1}{2} & Q_3(-Q_1 + \frac{1}{2}) \\
		-Q_1 - \frac{1}{2} & -Q_1 + Q_2 - \frac{1}{2} & Q_3(Q_1 + \frac{1}{2})
	\end{bmatrix},
\end{equation}
where $Q_1 = \frac{E}{2Q}$, $Q_2 = \frac{F}{k_1 Q}$, $Q_3 = \frac{k_3 y_B}{k_1}$.

The eigenvalue $ \lambda_1 = 0 $ indicates that disturbances remain unchanged along the direction perpendicular to the plane defined by $ \delta\psi_1 = \delta y_A + \delta y_B + \delta y_C = \text{Const} $. As a result, the reaction trajectory is orthogonal to the vector $[1,1,1]$ and confined to a two-dimensional plane in phase space. This invariance of $ \psi_1 = y_A + y_B + y_C $ reflects element conservation in the Robertson system.

The central panel of Figure~\ref{fig:robertson} shows the evolution of the second and third eigenvalues. The eigenvalue $ \lambda_2 $, associated with the fast reaction direction, increases continuously, indicating a decreasing characteristic time scale that ultimately reaches $ 10^{-4} $. The concentration $ y_B $, primarily influenced by rapid changes, reaches its initial equilibrium around $ 10^{-3} $. This suggests that perturbations in the fast direction dissipate quickly, establishing a quasi-steady state early in the reaction.

The eigenvalue $ \lambda_3 $ corresponds to the slow time scale $ \tau_3 $, which grows to approximately $ 10^7 $ as the system nears steady state. The bottom panel of Figure~\ref{fig:robertson} shows that $ \delta\psi_2 $ and $ \delta\psi_3 $ steadily decay, leaving only information associated with $ \delta\psi_1 $ after a certain period.

\subsubsection{Inference on the initial composition}

The ground truth of the initial composition $\boldsymbol{y}_0$ is assumed to follow a Gaussian distribution:
$\boldsymbol{\tilde{y}}_0 = \frac{\boldsymbol{y}_0 - \boldsymbol{\mu}_0}{\boldsymbol{s}_0} \sim \mathcal{N}(\boldsymbol{0}, \boldsymbol{I}),$
where the mean and standard deviation vectors are given by
$\boldsymbol{\mu}_0 = [0.95, 5 \times 10^{-6}, 0.05]$, $
    \boldsymbol{s}_0 = [0.01, 10^{-6}, 0.01]$.
The components of $\boldsymbol{y}_0$ are assumed to be independent.  

The target acceptance rate of the NUTS sampler is set to 0.8. The sequence of observation times $t_\mathrm{obs}$ is defined as $\left\{10^{-4.0}, 10^{-3.9}, 10^{-3.8}, \dots, 10^{0.0}\right\}$,
resulting in a total of 41 Bayesian inference experiments. The parameters $\boldsymbol{\mu}_\mathrm{obs}$ and $\boldsymbol{s}_\mathrm{obs}$ are obtained by integrating samples of $\boldsymbol{y}_0$ drawn from $\mathcal{N}(\boldsymbol{\mu}_0, \boldsymbol{s}_0)$ up to $t_\mathrm{obs}$.  
The prior distribution is defined as $\boldsymbol{\tilde{y}}_0 = \frac{\boldsymbol{y}_0 - \boldsymbol{\mu}_0}{\boldsymbol{s}_0} \sim \mathcal{U}(-3\cdot\boldsymbol 1, 3\cdot\boldsymbol 1)$, indicating that $\boldsymbol{y}_0$ bounded within an interval $\boldsymbol{\mu}_0 - 3\boldsymbol{s}_0 \leq \boldsymbol{y}_0 \leq \boldsymbol{\mu}_0 + 3\boldsymbol{s}_0.$

\begin{figure}[h]
	\centering
	\subfigure[$t_\mathrm{obs} = 10^{-4}$]{
			\includegraphics[width=0.3\textwidth]{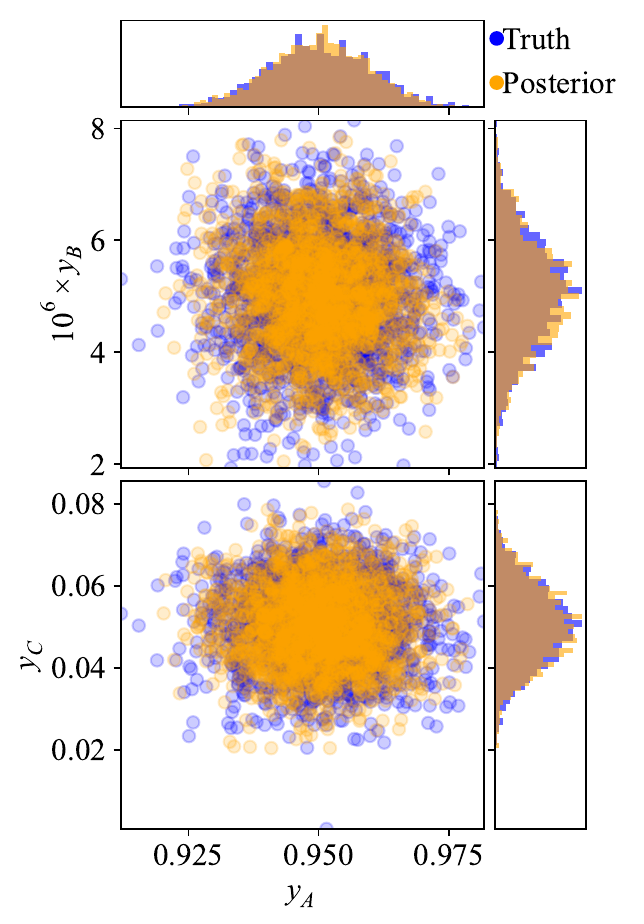}
			\label{fig:obsat1e-4}
		}
	\subfigure[$t_\mathrm{obs} = 1$]{
			\includegraphics[width=0.3\textwidth]{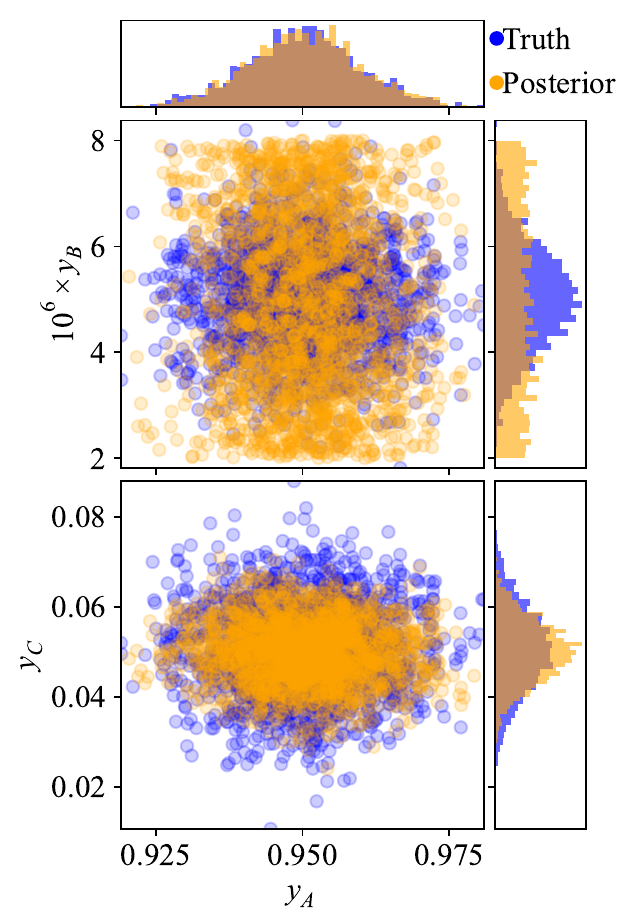} 
			\label{fig:obsat1}
		}
	\subfigure[Distributions evolved to $t = 1$]{
		\includegraphics[width=0.3\textwidth]{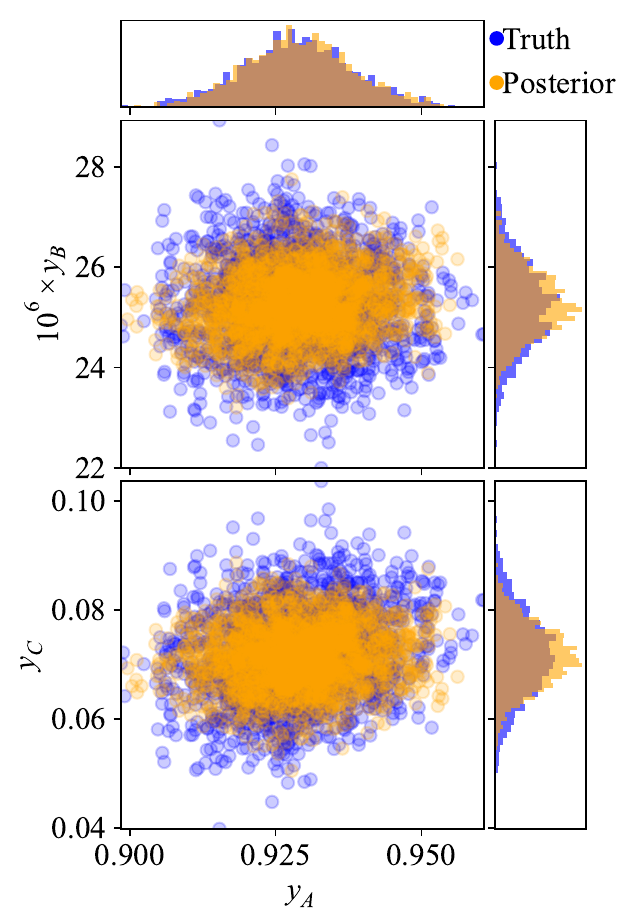} 
		\label{fig:predat1}
	}
	\caption[Inferenced initial values at different observation times]{Inference results for initial values at different observation times: (a) $t_\mathrm{obs} = 10^{-4}$, (b) $t_\mathrm{obs} = 1$. (c) shows the distributions obtained by evolving each sample in (b) to $ t = 1 $.}
	\label{fig:inference_2_results}
\end{figure}

Figure~\ref{fig:inference_2_results} presents the inference results at $t_\mathrm{obs} = 1$ and $10^{-4}$. When $t_\mathrm{obs}$ is close to zero, the posterior distribution closely aligns with the ground truth. However, as $t_\mathrm{obs} \gg 0$, the inference accuracy of $y_B$ decreases. Due to the rapid decay of certain components in a stiff system, some information is lost at $t_\mathrm{obs}$, leading to reduced inference accuracy for $y_B$.  

For instance, species $B$ is an intermediate product in the system, and its variations are governed by short time scales. Before $y_B$ reaches equilibrium, it rapidly attains a quasi-steady state (QSS). If QSS is already established at $t_\mathrm{obs}$, the initial value $y_{B0}$ has minimal influence on $y_B(t_\mathrm{obs})$, indicating that most of the information contained in $y_{B0}$ has been lost by $t_\mathrm{obs}$. As demonstrated in Figures~\ref{fig:obsat1} and \ref{fig:predat1}, although $y_B$ is uniformly distributed at the initial time, its distribution at the observation time does not significantly deviate from the true distribution, further supporting this observation.

The normalized posterior distribution is defined as:
\begin{equation}
	\tilde{P}_\mathrm{post}(\boldsymbol{\tilde{y}}_0) = {P}_\mathrm{post}(\boldsymbol{s}_0 \odot \tilde{\boldsymbol{y}}_0 + \boldsymbol{\mu}_0),
\end{equation}
where $\odot$ denotes Hadamard product.
Thus, the Jensen-Shannon divergence (JSD) between the ground truth and the posterior distribution is given by:
\begin{equation}
	D_\mathrm{JS}(P_\mathrm{truth} \parallel P_\mathrm{post}) = D_\mathrm{JS}(\mathcal{N}(\boldsymbol{0},\boldsymbol{I}) \parallel \tilde{P}_\mathrm{post}),
\end{equation}
and the JSD between the prior and posterior distributions is given by:
\begin{equation}
	D_\mathrm{JS}(P_\mathrm{prior} \parallel P_\mathrm{post}) = D_\mathrm{JS}(\mathcal{U}(-3\cdot\boldsymbol{1},3\cdot\boldsymbol{1}) \parallel \tilde{P}_\mathrm{post}).
\end{equation}

\begin{figure}[htpb]
	\centering
	\includegraphics[width=0.5\linewidth]{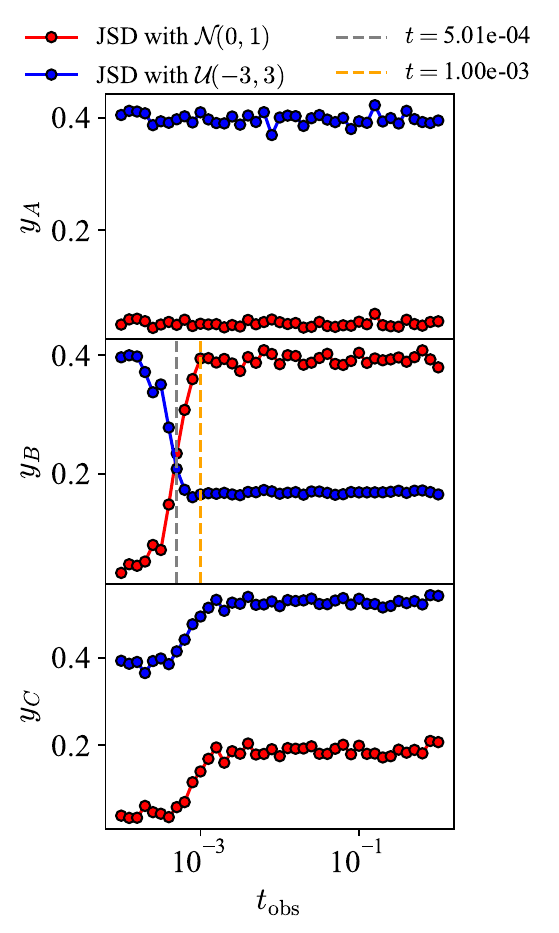}
	\caption{Variation of JSD with different $t_\mathrm{obs}$ without covariance conditions. 
		Red line: JSD between posterior and ground truth. 
		Blue line: JSD between posterior and prior. 
		Gray dashed line: the time when JSD with ground truth surpasses JSD with prior for the first time. 
		Orange dashed line: the time when inference failure occurs.}
	\label{fig:jsd1}
\end{figure} 


The variations of JSD with respect to $t_\mathrm{obs}$ are presented in Figure~\ref{fig:jsd1}. The JSD for $y_A$ remains approximately 0.03, indicating high inference accuracy. The normalized posterior $\tilde{P}_\mathrm{post}(\boldsymbol{\tilde{y}}_0)$ closely approximates $\mathcal{N}(\boldsymbol{0},\boldsymbol{I})$ at small $t_\mathrm{obs}$. However, as $t_\mathrm{obs}$ increases, the posterior distribution of $y_B$ gradually converges toward a uniform distribution.

By analyzing the posterior and prior distributions of $y_B$ at each $t_\mathrm{obs}$, we establish the following criterion for inference failure:
\begin{equation}
D_\mathrm{JS}(P_\mathrm{truth} \parallel P_\mathrm{post}) - D_\mathrm{JS}(P_\mathrm{prior} \parallel P_\mathrm{post}) > 0.2.
\label{eq:fail_judge}
\end{equation}
If this condition is met, the difference between the posterior and prior distributions becomes indistinguishable. Equation~\eqref{eq:fail_judge} is thus used as the criterion for inference failure in this study.

The JSD values between the posterior and ground truth distributions for $ y_B $ and $ y_C $ increase notably around $ t_\mathrm{obs} = 5 \times 10^{-4} $. However, their posterior distributions behave differently. As shown in Figure~\ref{fig:obsat1}, the posterior of $ y_B $ approaches a uniform distribution, aligning with the prior, indicating inference failure due to the algorithm's inability to distinguish between the posterior and prior. In contrast, the standard deviation of $ y_C $'s posterior decreases slightly, while its mean remains consistent with the ground truth (see Figure~\ref{fig:predat1}).

\subsubsection{Significance of the composition covariance matrix}

\begin{figure}[htpb]
	\centering
	\includegraphics[width=0.5\linewidth]{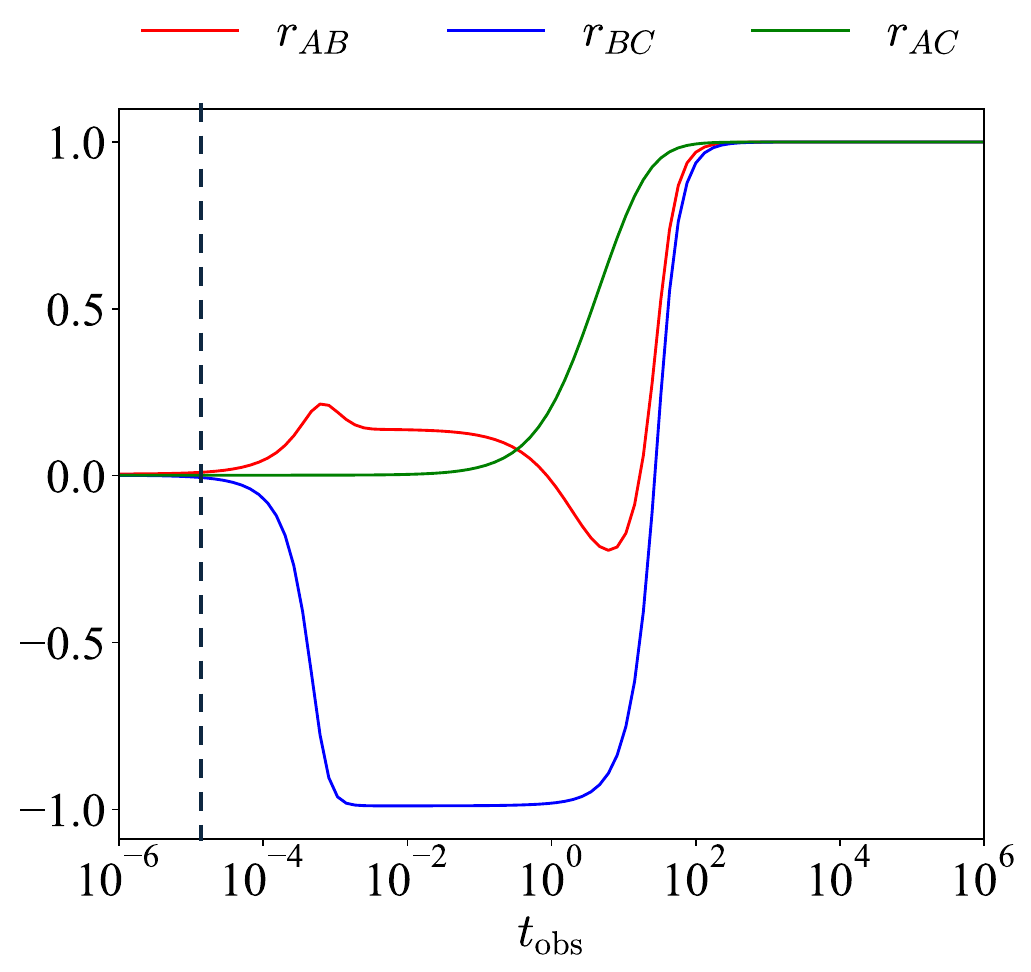}
	\caption{Correlation coefficients between $y_A$, $y_B$, and $y_C$. The black dashed line splits the two stages of whether the three components are independent or not.}
	\label{fig:r_plot}
\end{figure} 

A portion of the inference error stems from information loss along characteristic directions associated with shorter time scales, leading to the inference failure observed in Figure~\ref{fig:obsat1}. Additionally, the Gaussian likelihood function assumes a diagonal covariance matrix, implying independence among $ y_A $, $ y_B $, and $ y_C $ at observation times. However, as shown in Figure~\ref{fig:r_plot}, this independence indicated by the close-to-zero correlation coefficients holds only shortly after the reaction begins.

To incorporate the correlation between $y_A$, $y_B$, and $y_C$, we replace $\boldsymbol{s}_\mathrm{obs}$ in Equation~\eqref{eq:posterior} with the full covariance matrix $\boldsymbol{\Sigma}_\mathrm{obs}$, leading to the modified posterior distribution:
\begin{equation}
	P_{\mathrm{post}}(\boldsymbol{y}_0 | \boldsymbol{\mu}_{\mathrm{obs}}, \boldsymbol{\Sigma}_{\mathrm{obs}}) 
	\propto \exp \left( -\frac{1}{2} (\boldsymbol{y}(t_{\mathrm{obs}}) - \boldsymbol{\mu}_{\mathrm{obs}})^T \boldsymbol{\Sigma}_{\mathrm{obs}}^{-1} (\boldsymbol{y}(t_{\mathrm{obs}}) - \boldsymbol{\mu}_{\mathrm{obs}}) \right).
\end{equation}

\begin{figure}[htpb]
	\centering
	\includegraphics[width=0.5\linewidth]{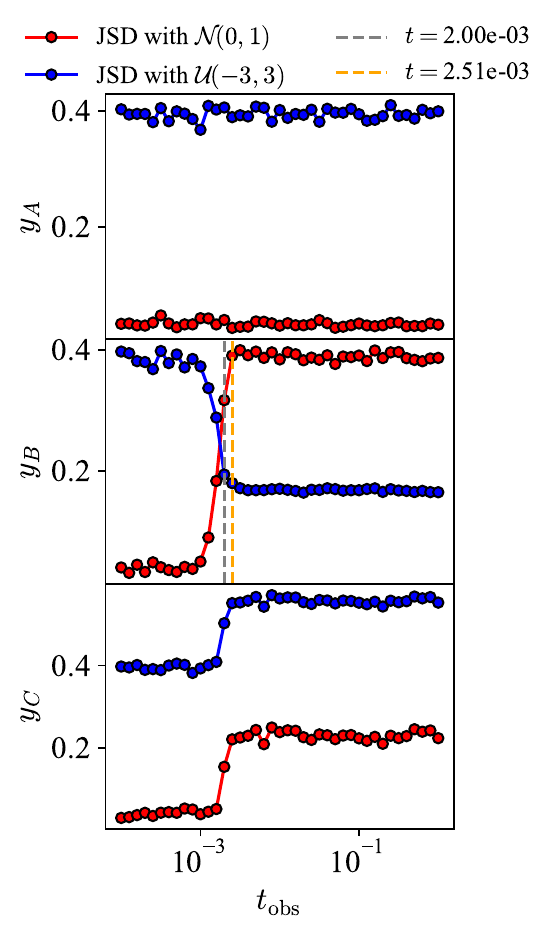}
	\caption{Variation of JSD with different $t_\mathrm{obs}$ with covariance conditions. 
		Red line: JSD between posterior and ground truth. 
		Blue line: JSD between posterior and prior. 
		Gray dashed line: the time when JSD with ground truth surpasses JSD with prior for the first time. 
		Orange dashed line: the time when inference failure occurs.}
	\label{fig:jsd2}
\end{figure}

With this modification, Bayesian inference was performed using the same configuration as before. The JSD values after incorporating the covariance matrix are shown in Figure~\ref{fig:jsd2}. The inference failure time of $ y_B $ ($t=2.51\times 10^{-3}$) occurs later than in the case without covariance ($t=10^{-3}$), supporting the hypothesis that part of the inference error arises from ignoring correlations among $ y_A $, $ y_B $, and $ y_C $ at the observation times.

\subsubsection{Connection between information loss and rank deficiency in the sensitivity matrix}
\label{sec:mappcov}

\begin{figure}[htbp]
	\centering
	\includegraphics[width=0.5\linewidth]{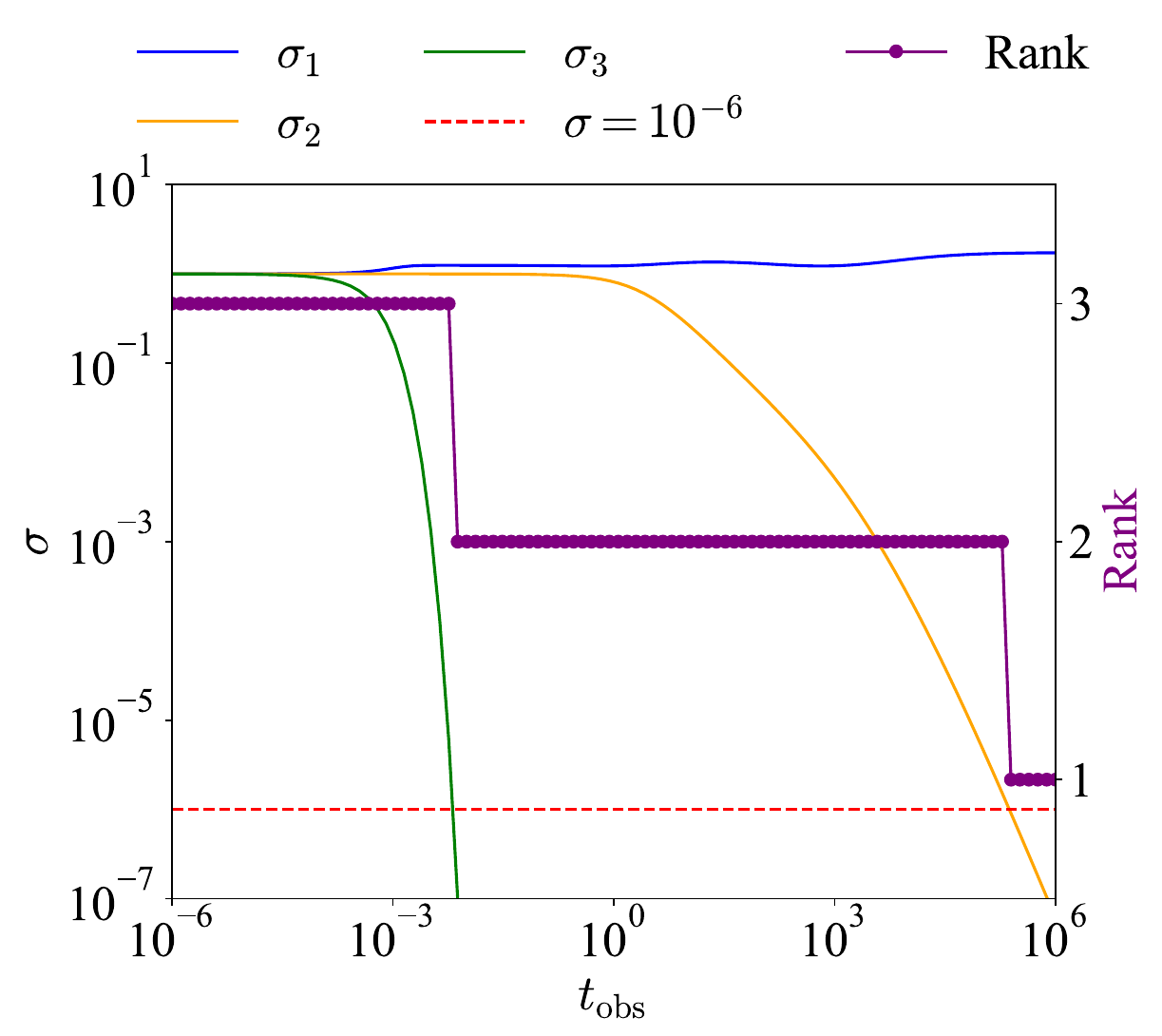}
	\caption{The rank and singular values of the sensitivity matrix \( \boldsymbol{A} \) as a function of \( t_\mathrm{obs} \). The purple points represent the rank of \( \boldsymbol{A} \), while the solid lines represent its singular values. }
	\label{fig:rank_A_rob}
\end{figure}

Uncertainty propagates from the initial state to the observed quantities through the sensitivity matrix \( \boldsymbol{A} \). The covariance of the observed quantities is given by:

\begin{equation}
	\boldsymbol{\Sigma}_\mathrm{obs} = \boldsymbol{A} \boldsymbol{\Sigma}_0 \boldsymbol{A}^T.
\end{equation}

When inferring the initial uncertainty, the posterior estimate of \( \boldsymbol{\Sigma}_0 \) is obtained as:

\begin{equation}
	\boldsymbol{\Sigma}_0 = \boldsymbol{A}^{-1} \boldsymbol{\Sigma}_\mathrm{obs} \boldsymbol{A}^{-T}.
\end{equation}
For a detailed derivation of the uncertainty propagation, see~\ref{appendix:uncertainty_propagation}. 

As shown in Figure~\ref{fig:rank_A_rob}, the rank of sensitivity matrix \( \boldsymbol{A} \) is decreasing as $t_\mathrm{obs}$ increasing, making \( \boldsymbol{A} \) not always invertible. As the system evolves, local equilibrium emerges, and the reaction trajectory collapses onto a lower-dimensional manifold, making initial composition inference challenging. The loss of rank corresponds to a loss of information along certain characteristic directions, leading to inference failure. Furthermore, the time when the rank drops for the first time and the time when inference fails are both in the range of $10^{-3}$ to $10^{-2}$.

\subsection{Hydrogen autoignition system ($ n_e = 3 $, $ n_s = 9 $, $ n_r = 28 $)}
To further validate the methodology, we apply the framework to the \href{https://github.com/DENG-MIT/Arrhenius.jl/blob/main/mechanism/h2o2.yaml}{hydrogen-oxygen mechanism}, which consists of the following species:
$\mathrm{H}_2$, $\mathrm{H}$, $\mathrm{O}$, $\mathrm{O}_2$, $\mathrm{OH}$, $\mathrm{H}_2$, $\mathrm{O}$, $\mathrm{HO}_2$, $\mathrm{H}_2$, $\mathrm{O}_2$, and $\mathrm{N}_2$.
The ground truth of the initial mass fractions $\boldsymbol{\phi}_0 \in \mathbb{R}^9$ (following the order of species enumeration) is assumed to follow a normal distribution: $\boldsymbol{\phi}_0 \sim \mathcal{N}(\boldsymbol{\mu}_0, \boldsymbol{s}_0)$,
where $\boldsymbol{\mu}_0 = [0.1, 10^{-3}, 10^{-3}, 0.2, 10^{-3}, 10^{-3}, 10^{-3}, 10^{-3}, 0.694]$, and $\boldsymbol{s}_0 = \boldsymbol{\mu}_0 / 10$.  
The homogeneous, insulated, and isobaric reactor is described in Equation~\eqref{eq:ODE_formula}. The initial pressure is set to $P_0 = 1 \ \mathrm{atm}$, and all initial conditions share the same initial enthalpy $h_0$, which is calculated at $T = 1200\ \mathrm{K}, P = P_0, \boldsymbol{\phi} = \boldsymbol{\mu}_0$.  
The posterior distribution is given by:
\begin{equation}
	P_\mathrm{post}(\boldsymbol{\phi}_0 | \boldsymbol{\mu}_\mathrm{obs}, \boldsymbol{\Sigma}_\mathrm{obs}) 
	\propto \exp \left( -\frac{1}{2} (\boldsymbol{R}(\boldsymbol{\phi}_0 ,t_\mathrm{obs}) - \boldsymbol{\mu}_\mathrm{obs})^T 
	\boldsymbol{\Sigma}_\mathrm{obs}^{-1} (\boldsymbol{R}(\boldsymbol{\phi}_0 ,t_\mathrm{obs}) - \boldsymbol{\mu}_\mathrm{obs}) \right).
\end{equation}
If the likelihood function employs the standard deviation $\boldsymbol{s}_0$, the off-diagonal elements of the covariance matrix $\boldsymbol{\Sigma}_0$ can be set to 0. The sequence of observation times is defined as  
$\{10^{-8.0}$,$ 10^{-7.95}$,$ 10^{-7.90}$,$ \dots$,$ 10^{-1.0}$$ \}$ (s).

\subsubsection{Inference on the initial composition}

\begin{figure}[htb]
	\centering
	\subfigure[Species with successful inference]{
		\includegraphics[width=0.5\textwidth]{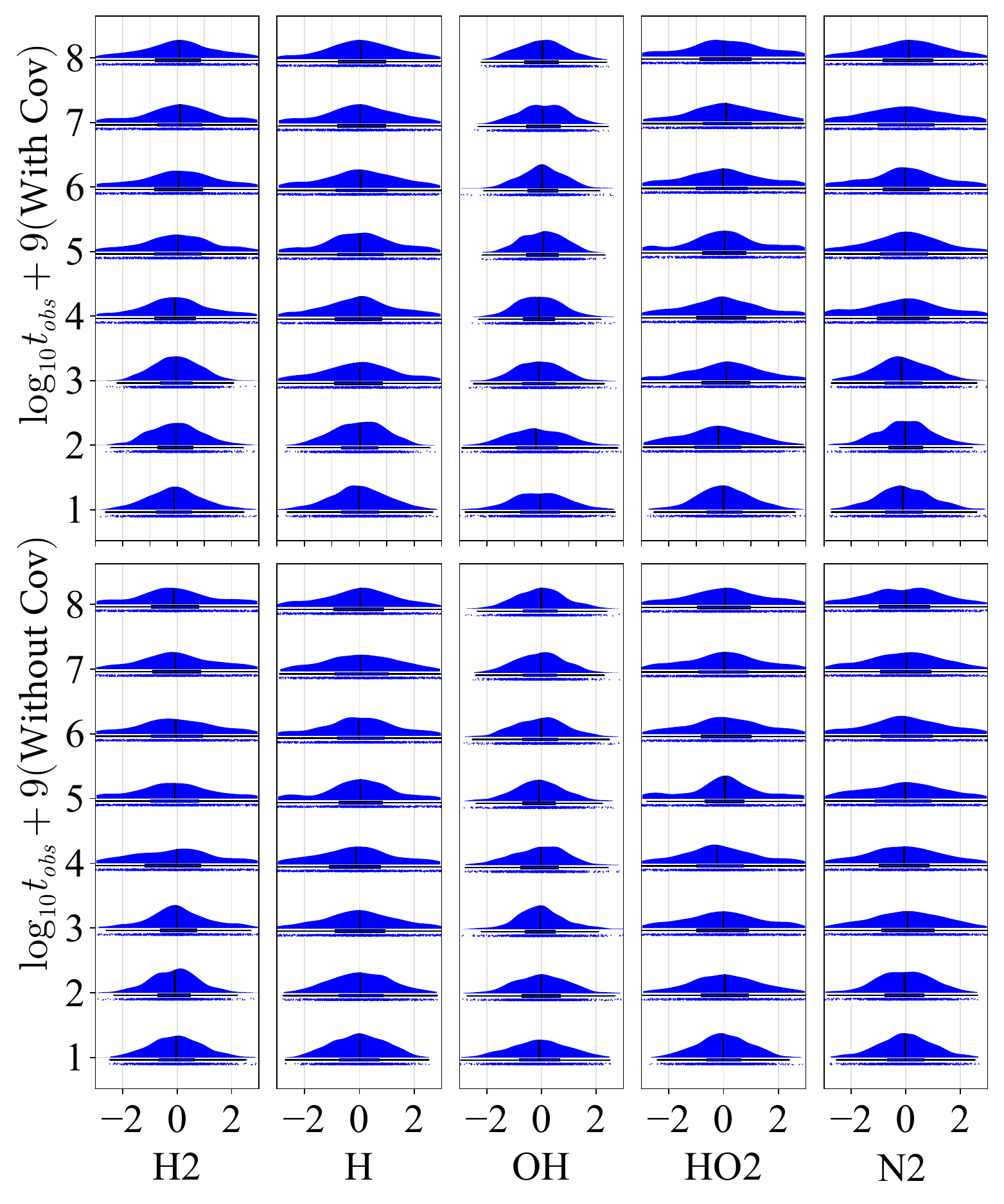}
		\label{fig:success}
	}
	\subfigure[Species with inference failure]{
		\includegraphics[width=0.4\textwidth]{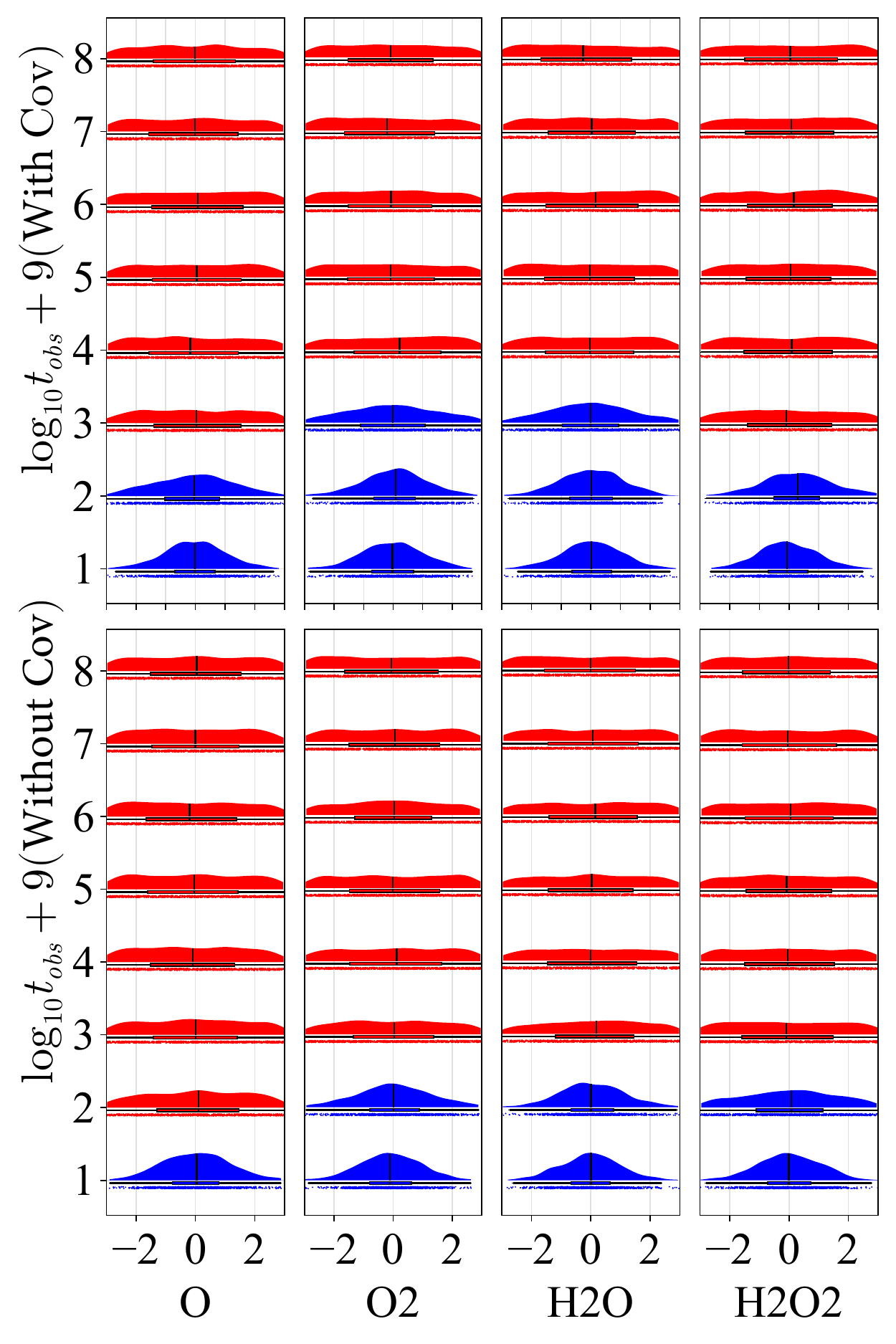} 
		\label{fig:fail}
	}
	\caption[Raincloud plot of the posterior]{Raincloud plots of the posterior distributions. (a) shows species whose inference is always successful (blue rainclouds), while (b) highlights species where inference failure occurs (red rainclouds).}
	\label{fig:rain_cloud}
\end{figure}

The normalized posteriors $\tilde{P}_\mathrm{post}(\tilde{\boldsymbol{\phi}}_0) = P_\mathrm{post}(\boldsymbol{s} \odot \tilde{\boldsymbol{\phi}}_0 + \boldsymbol{\mu})$ for $t_\mathrm{obs} = 10^{-8}, 10^{-7}, \dots, 10^{-1}$ (s) are shown in Figure~\ref{fig:rain_cloud}. The raincloud plots satisfying Equation~\eqref{eq:fail_judge} are highlighted in red. Figure~\ref{fig:rain_cloud} demonstrates that incorporating species correlation information can improve inference accuracy under certain observation times $t_\mathrm{obs}$. For instance, for species O, inference using variance at $t_\mathrm{obs} = 10^{-7} \ \mathrm{s}$ fails, whereas incorporating covariance information leads to successful inference. 

\begin{table}[htbp]
	\centering
	\begin{tabular}{lcccc}
		\toprule
		& \ch{O} & \ch{O2} & \ch{H2O} & \ch{H2O2} \\
		\midrule
		Covariance & $10^{-6.75}$ & $10^{-5.6}$ & $10^{-5.05}$ & $10^{-6.65}$ \\
		Variance   & $10^{-7.0}$  & $10^{-6.1}$ & $10^{-6.0}$  & $10^{-6.95}$ \\
		\bottomrule
	\end{tabular}
	\caption{Critical failure times (s) for different species under covariance and variance-based inference.}
	\label{tab:failure_times}
\end{table}

For each $t_\mathrm{obs}$, Equation~\eqref{eq:fail_judge} was used to determine whether the inference failed. The critical failure times of inference for different species are presented in Table~\ref{tab:failure_times}. Inferences supplemented with covariance information remain valid for a longer observation period before failing, further emphasizing the necessity of incorporating covariance information in initial value inference.

\subsubsection{Causes of rank reduction in sensitivity matrix of stiff reactive system}
\label{sec:species_times}

In Section~\ref{sec:mappcov}, we established that the covariance mapping \(\boldsymbol{\Sigma}_\mathrm{obs} = \boldsymbol{A} \boldsymbol{\Sigma}_0 \boldsymbol{A}^{T}\) holds in the hydrogen/oxygen system. When \(\boldsymbol{A}\) is singular, \(\operatorname{rank}(\boldsymbol{\Sigma}_\mathrm{obs}) < \operatorname{rank}(\boldsymbol{\Sigma}_0)\), indicating covariance loss during evolution, which ultimately causes inference failure.

\begin{figure}[htpb]
	\centering
	\includegraphics[width=0.5\linewidth]{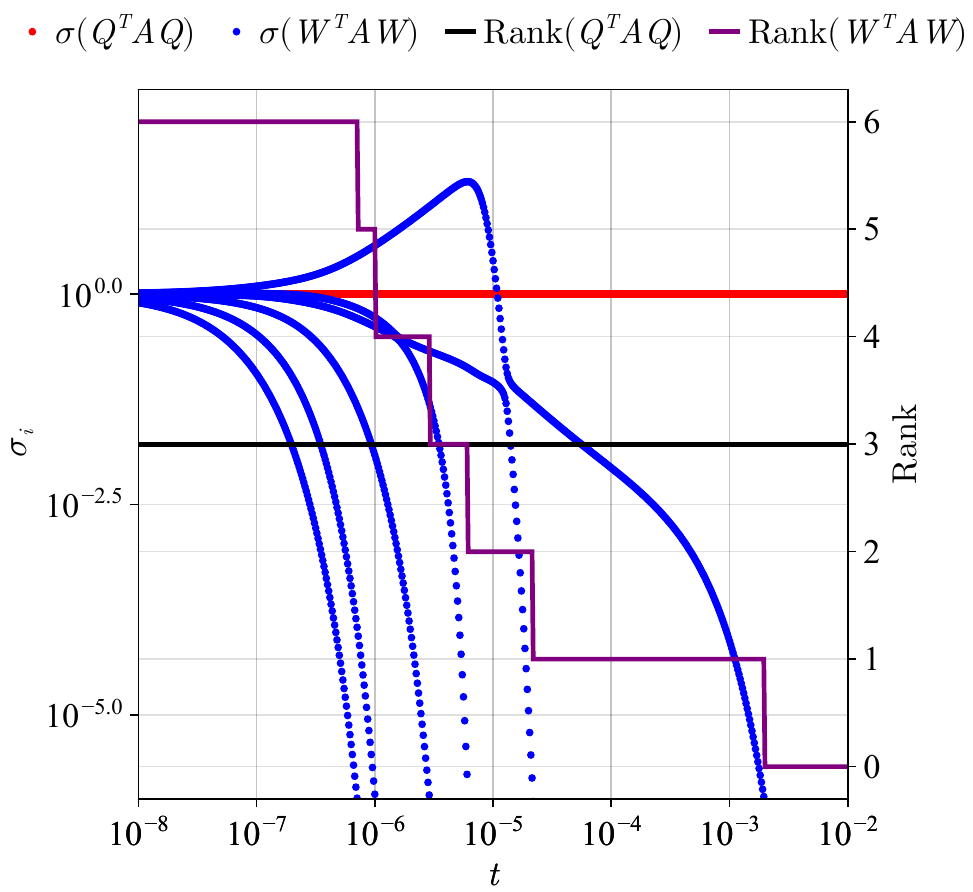}
	\caption{Singular value and rank evolution of $\boldsymbol{Q}^T \boldsymbol{A} \boldsymbol{Q}$ and $\boldsymbol{W}^T \boldsymbol{A} \boldsymbol{W}$ over time $t$ (s). Red dots: singular values of $\boldsymbol{Q}^{T} \boldsymbol{A} \boldsymbol{Q}$, blue dots: singular values of $\boldsymbol{W}^T \boldsymbol{A} \boldsymbol{W}$, purple solid line: rank of $\boldsymbol{W}^T \boldsymbol{A} \boldsymbol{W}$, black solid line: rank of $\boldsymbol{Q}^{T} \boldsymbol{A} \boldsymbol{Q}$. The rank of $\boldsymbol{Q}^{T} \boldsymbol{A} \boldsymbol{Q}$ stays at 3, while $\boldsymbol{W}^T \boldsymbol{A} \boldsymbol{W}$ gradually drops to 0.}
	\label{fig:rank_a_waq_waw_h2o2}
\end{figure}

The sensitivity matrix \( \boldsymbol{A} \) governs the propagation of perturbations from initial conditions to observed states. To analyze information loss, we decompose the system into two subspaces:

\begin{itemize}
    \item \textbf{Conserved Space}: Defined by element conservation, this space is spanned by a column-orthogonal matrix \( \boldsymbol{Q} \in \mathbb{R}^{n_s \times n_e} \), which represents constraints imposed by mass conservation.
    \item \textbf{Reaction Space (Orthogonal Complement)}: The remaining degrees of freedom are captured by \( \boldsymbol{W} \in \mathbb{R}^{n_s \times (n_s - n_e)} \), which spans the directions where chemical reactions actively evolve.
\end{itemize}

Perturbations in these subspaces follow the transformation:
\begin{equation}
        \label{eq:pert}
	\binom{d \boldsymbol{R}_q}{d \boldsymbol{R}_w}=
	\begin{pmatrix}
		\boldsymbol{I} & \mathbf{0} \\
		\boldsymbol{W}^T \boldsymbol{A} \boldsymbol{Q} & \boldsymbol{W}^T \boldsymbol{A} \boldsymbol{W}
	\end{pmatrix}
	\binom{d \boldsymbol{\phi}_{0 q}}{d \boldsymbol{\phi}_{0 w}}.
\end{equation}
Here, \( d\boldsymbol{R}_q=\boldsymbol{Q}^Td\boldsymbol{R}\) and \( d\boldsymbol{\phi}_{0q}=\boldsymbol{Q}^Td\boldsymbol{\phi}_{0}\) correspond to perturbations in the conserved space, which remain unaffected due to element conservation, while \( d\boldsymbol{R}_w=\boldsymbol{W}^Td\boldsymbol{R}\) and \(d\boldsymbol{\phi}_{0w}=\boldsymbol{W}^Td\boldsymbol{\phi}_{0}\) represents perturbations in the reaction space. The evolution of \( d\boldsymbol{R}_w \) is given by:
\begin{equation}
	d \boldsymbol{R}_w = \boldsymbol{W}^T \boldsymbol{A} \boldsymbol{Q} d \boldsymbol{\phi}_{0 q} + \boldsymbol{W}^T \boldsymbol{A} \boldsymbol{W} d \boldsymbol{\phi}_{0 w},
	\label{eq:dyw}
\end{equation}
where:
\begin{equation}
	\lim_{t\to+\infty} \boldsymbol{W}^T \boldsymbol{A} \boldsymbol{W}=0.
\end{equation}
This implies that \( \boldsymbol{W}^T \boldsymbol{A} \boldsymbol{W} \) gradually loses rank, eventually reaching 0. Consequently, all information in the reaction space is lost, preventing the inference of initial perturbations \( d \boldsymbol{\phi}_{0 w} \) from observed data \( d \boldsymbol{R}_w \), leading to inference failure.For a complete derivation, see Appendix~\ref{appendix:reaction_space}.

Figure~\ref{fig:rank_a_waq_waw_h2o2} shows the rank evolution of \( \boldsymbol{Q}^{T} \boldsymbol{A} \boldsymbol{Q} \) and \( \boldsymbol{W}^T \boldsymbol{A} \boldsymbol{W} \). The rank of \( \boldsymbol{W}^T \boldsymbol{A} \boldsymbol{W} \) decreases over time, causing information loss in \( d \boldsymbol{R}_{w} \), while \( \boldsymbol{W}^T \boldsymbol{A} \boldsymbol{Q} \) remains unchanged. The critical failure times identified earlier align with the rank reduction of \( \boldsymbol{W}^T \boldsymbol{A} \boldsymbol{W} \), though the exact quantitative relationship remains uncertain.

Inference failure arises from both systemic and observational factors: system stiffness leads to rank deficiency and information loss, while correlations between observation times impact inference stability. Assessing system stiffness is crucial for evaluating inference reliability.

\subsection{Extension to non-adiabatic systems}
\label{sec:disc}
\begin{figure}[htpb]
	\centering
	\includegraphics[width=0.5\linewidth]{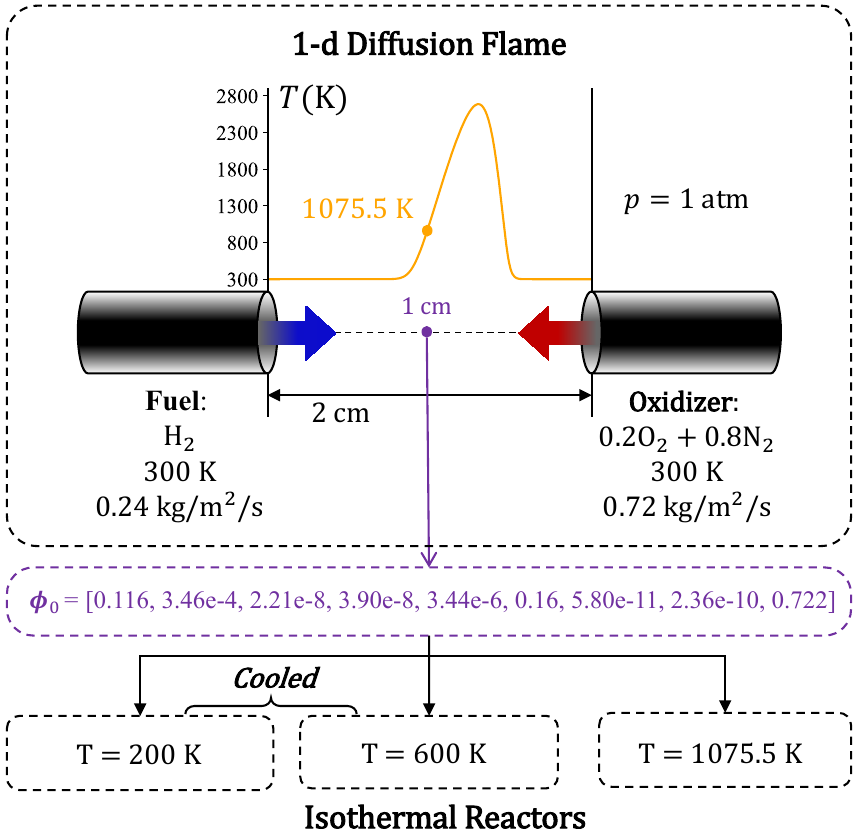}
	\caption{Schematic of the one-dimensional diffusion flame. The fuel (\ce{H2}) and oxidizer (\ce{O2} + \ce{Ar}) are injected from opposite sides at 300 K with mass fluxes of 0.24 kg/m$^2$/s and 0.72 kg/m$^2$/s, respectively, at a pressure of 1 atm. The sample is taken at $x = 1$ cm, where the initial composition mean is defined as $\boldsymbol\phi_0$, then it flows into three isothermal reactors at different temperatures, each simulating a different level of cooling. The temperature profile along the flame axis is also shown.}
	\label{fig:diffusion1d}
\end{figure}

Real combustion systems often experience heat loss during the reacting process. Therefore, we apply the method to a simplified probe sampling case to illustrate the effect of a quenching device at the probe tip.Figure~\ref{fig:diffusion1d} depicts the one-dimensional diffusion flame scenario adopted in this section. A probe samples gas at $x = 1 \, \text{cm}$, and the sampled gas is subjected to three quenching treatments: rapidly cooled to 200 K or 600 K, and without cooling (1075.5 K). The sample gas remains at the cooled temperature inside the probe, ensuring that $T$ remains constant over time in the stiff ODE system.

\begin{figure}[tbp]
	\centering
	\includegraphics[width=1.0\linewidth]{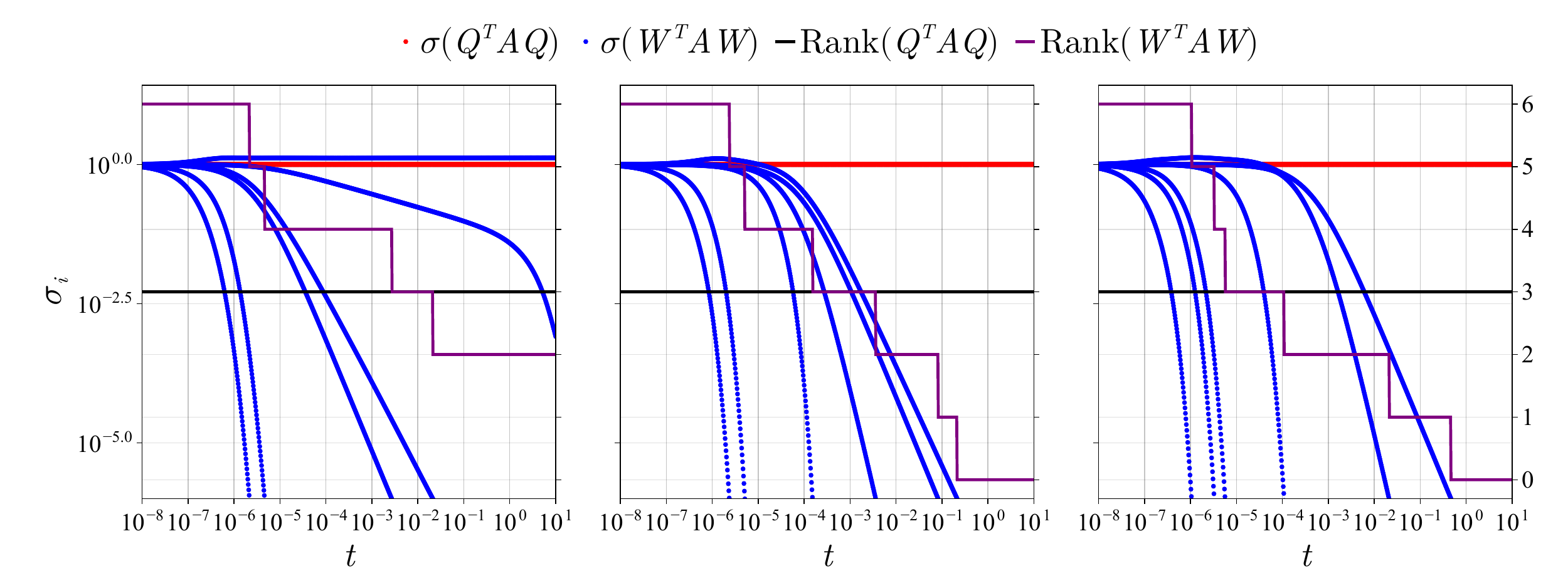}
	\caption{Singular values and ranks of $\boldsymbol{Q}^T \boldsymbol{A} \boldsymbol{Q}$ and $\boldsymbol{W}^T \boldsymbol{A} \boldsymbol{W}$ over time $t$ (s). From left to right: cooled to 200 K; cooled to 600 K; without cooling.Red dots: singular values of $\boldsymbol{Q}^{T} \boldsymbol{A} \boldsymbol{Q}$, blue dots: singular values of $\boldsymbol{W}^T \boldsymbol{A} \boldsymbol{W}$, purple solid line: rank of $\boldsymbol{W}^T \boldsymbol{A} \boldsymbol{W}$, black solid line: rank of $\boldsymbol{Q}^{T} \boldsymbol{A} \boldsymbol{Q}$.}
	\label{fig:cooling_rank_descent}
\end{figure}

\begin{table}[tbp]
	\centering
	\caption{Rank descent times (s) under different cooling conditions.}
	\begin{tabular}{lcccccc}
		\toprule
		\textbf{Temperature} & \textbf{1st} & \textbf{2nd} & \textbf{3rd} & \textbf{4th} & \textbf{5th} & \textbf{6th} \\
		\midrule
		1075.5 K & $10^{-5.98}$ & $10^{-5.49}$ & $10^{-5.25}$ & $10^{-3.97}$ & $10^{-1.68}$ & $10^{-0.34}$  \\
		600 K & $10^{-5.63}$ & $10^{-5.30}$ & $10^{-3.82}$ & $10^{-2.45}$ & $10^{-1.09}$ & $10^{-0.68}$ \\
		200 K & $10^{-5.67}$ & $10^{-5.34}$ & $10^{-2.57}$ & $10^{-1.68}$ & \textemdash & \textemdash  \\
		\bottomrule
	\end{tabular}
	\label{tab:cooling_rank_descent}
\end{table}

The same analysis as in Section~\ref{sec:species_times} is performed for each case. The rank descent times are listed in Table~\ref{tab:cooling_rank_descent}, while singular values and ranks as functions of time are shown in Figure~\ref{fig:cooling_rank_descent}. The singular values of $\boldsymbol{W}^T \boldsymbol{A} \boldsymbol{W}$, denoted as $\sigma_1, \sigma_2, \ldots, \sigma_6$ in descending order, reveal the following trends. The first and second rank descents exhibit minimal variation with temperature, indicating that $\sigma_1$ and $\sigma_2$ are relatively insensitive to temperature changes. Conversely, the descent rates of $\sigma_3$ and $\sigma_4$ slow down as temperature decreases, leading to an increase in rank descent times. This suggests that the quenching device mitigates information loss to some extent, particularly under low-temperature conditions (200 K). Notably, $\sigma_5$ does not fall below the threshold within the observation period, while $\sigma_6$ remains constant.

\newpage

\section{Conclusions}
\label{sec:Ccl}

This study establishes two key conclusions regarding initial composition inference in stiff reactive systems. First, inference failure is directly linked to the dimension loss of the sensitivity matrix, which eventually leads to complete information loss within the reaction space while preserving information within the conserved space. Second, incorporating species correlation information, such as observational covariance, mitigates the issue of inference failure. Specifically, in Bayesian inference for hydrogen autoignition systems, the introduction of covariance can delay the critical failure time of failing species by an order of magnitude, as seen for species O, from \(10^{-7}~\mathrm{s}\) to \(10^{-6}~\mathrm{s}\).

To quantify inference failure and information loss, two criteria are defined: a threshold-based Jensen-Shannon distance difference for determining inference failure at the species level and numerical rank reduction of the sensitivity matrix. While both methods quantify the extent of information loss, the former provides species-specific failure times but incurs higher computational costs.

In practical probe sampling, quenching helps preserve sample gas information. However, as demonstrated in the hydrogen diffusion flame case, information along two eigen-directions associated with fast reactions decays within \(10^{-5}~\mathrm{s}\) even at a fixed temperature of \(200\ \mathrm{K}\). For probe sampling, the impact of the decay of information along fast directions on species should be carefully quantified even at high quenching rate.

\section*{Acknowledgement}
The work was supported by National Natural Science Foundation of China (52106166).

\newpage
\appendix

\section{Uncertainty propagation in Bayesian inference}
\label{appendix:uncertainty_propagation}

For a normally distributed initial state \( \boldsymbol{y}_0 \) with mean \( \boldsymbol{\mu}_0 \) and covariance \( \boldsymbol{\Sigma}_0 \), it can be expressed in terms of a standard normal random vector \( \boldsymbol{z} \):
\begin{equation}
\label{eq:aaa}
	\boldsymbol{y}_0 = \boldsymbol{\mu}_0 + \boldsymbol{K} \boldsymbol{z},
\end{equation}
where \( \boldsymbol{K} \) satisfies \( \boldsymbol{K} \boldsymbol{K}^T = \boldsymbol{\Sigma}_0 \), which can be obtained via Cholesky decomposition.
For small perturbations, the system response at the observation time \( t_{\text{obs}} \) follows a linearized form:
\begin{equation}
	\boldsymbol{R}(\boldsymbol{y}_0, \boldsymbol{t}_{\text{obs}}) = \boldsymbol{R}(\boldsymbol{\mu}_0, \boldsymbol{t}_{\text{obs}}) + \boldsymbol{A} (\boldsymbol{y}_0 - \boldsymbol{\mu}_0).
\end{equation}
Substituting Equation~\eqref{eq:aaa} we obtain:
\begin{equation}
	\boldsymbol{R}(\boldsymbol{y}_0, \boldsymbol{t}_{\text{obs}}) = \boldsymbol{R}(\boldsymbol{\mu}_0, \boldsymbol{t}_{\text{obs}}) + \boldsymbol{A} \boldsymbol{K} \boldsymbol{z}.
\end{equation}
Thus, the covariance of the observed quantities is:
\begin{equation}
	\boldsymbol{\Sigma}_\mathrm{obs} = \boldsymbol{A} \boldsymbol{\Sigma}_0 \boldsymbol{A}^T.
\end{equation}
In Bayesian inference, the initial covariance is estimated from the observed quantities as:
\begin{equation}
	\boldsymbol{\Sigma}_0 = \boldsymbol{A}^{-1} \boldsymbol{\Sigma}_\mathrm{obs} \boldsymbol{A}^{-T}.
\end{equation}

\section{Perturbation propagation in the conserved and reaction spaces}
\label{appendix:reaction_space}

According to element conservation, there exists a matrix \( \boldsymbol{C} \in \mathbb{R}^{n_s \times n_e} \) such that:
\begin{equation}
	\boldsymbol{C}^T \boldsymbol{R} = \text{Const}.
\end{equation}
Define \( \boldsymbol{Q} \in \mathbb{R}^{n_s \times n_e} \) as a column-orthogonal matrix spanning the same column space as \( \boldsymbol{C} \), and let \( \boldsymbol{W} \in \mathbb{R}^{n_s \times (n_s - n_e)} \) be its orthogonal complement. Any perturbation \( d \boldsymbol{R} \) can then be decomposed as:
\begin{equation}
	d \boldsymbol{R} = \boldsymbol{Q} d \boldsymbol{R}_q + \boldsymbol{W} d \boldsymbol{R}_w.
\end{equation}

Applying this decomposition to both the initial perturbation and the perturbation at the observation time yields:

\begin{equation}
	\boldsymbol{Q} d \boldsymbol{R}_q + \boldsymbol{W} d \boldsymbol{R}_w = \boldsymbol{A} (\boldsymbol{Q} d \boldsymbol{\phi}_{0 q} + \boldsymbol{W} d \boldsymbol{\phi}_{0 w}).
\end{equation}
Since element conservation enforces \( d \boldsymbol{R}_q \equiv d \boldsymbol{\phi}_{0 q} \), we obtain:
\begin{equation}
	\boldsymbol{W} d \boldsymbol{R}_w = (\boldsymbol{A} \boldsymbol{Q} - \boldsymbol{Q}) d \boldsymbol{\phi}_{0 q} + \boldsymbol{A} \boldsymbol{W} d \boldsymbol{\phi}_{0 w}.
\end{equation}
Multiplying both sides by \( \boldsymbol{W}^T \) gives:
\begin{equation}
	d \boldsymbol{R}_w = \boldsymbol{W}^T \boldsymbol{A} \boldsymbol{Q} d \boldsymbol{\phi}_{0 q} + \boldsymbol{W}^T \boldsymbol{A} \boldsymbol{W} d \boldsymbol{\phi}_{0 w}.
\end{equation}
Expanding $d\boldsymbol{R} = \boldsymbol{A} d\boldsymbol{\phi}_0$ in the \( \boldsymbol{Q} \)-\( \boldsymbol{W} \) basis, define \( \boldsymbol{G}=[\boldsymbol{Q}\ \boldsymbol{W}] \) and let \( \tilde{\boldsymbol A} = \boldsymbol{G}^T \boldsymbol{A} \boldsymbol{G} \), leading to:
\begin{equation}
    \begin{aligned}
        \tilde{\boldsymbol{A}} &= \begin{pmatrix}
		\boldsymbol{Q}^T \boldsymbol{A} \boldsymbol{Q} & \boldsymbol{Q}^T \boldsymbol{A} \boldsymbol{W} \\
		\boldsymbol{W}^T \boldsymbol{A} \boldsymbol{Q} & \boldsymbol{W}^T \boldsymbol{A} \boldsymbol{W}
	\end{pmatrix},\\
        \boldsymbol{G}^Td\boldsymbol{R} &= \tilde{\boldsymbol{A}}\boldsymbol{G}^Td\boldsymbol{\phi}_0
    \end{aligned}
\end{equation}
Since element conservation ensures:
\begin{equation}
	\boldsymbol{Q}^{T} \boldsymbol{A} \boldsymbol{Q}=\boldsymbol{I}, \quad \boldsymbol{Q}^{T} \boldsymbol{A} \boldsymbol{W}=\mathbf{0},
\end{equation}
hence we obtain Equation~\eqref{eq:pert}.

At equilibrium, the perturbations in the reaction space and their initial conditions become independent:

\begin{equation}
	 \lim_{t\to+\infty}\frac{\partial \boldsymbol{R}_w}{\partial \boldsymbol{\phi}_{0w}} =\lim_{t\to+\infty}\boldsymbol{W}^{T} \boldsymbol{A} \boldsymbol{W} = \mathbf{0}.
\end{equation}

This confirms that all information in the reaction space is lost as \( \boldsymbol{W}^T \boldsymbol{A} \boldsymbol{W} \) degenerates, leading to inference failure. 
As the reaction proceeds, the dimension of the feasible domain of $d\boldsymbol{R}$ gradually decreases to $n_e$, which means that $\boldsymbol{R}$ falls on the $n_e$-dimensional attracting manifold. The reaction space undergoes progressive information loss, while the conserved space retains complete information.





\newpage
\bibliographystyle{elsarticle-num-names}
\bibliography{refs}

%
%
%
\end{document}